\documentclass[letterpaper, 10 pt, conference]{ieeeconf} 

\IEEEoverridecommandlockouts

\usepackage{mathptmx}
\usepackage{amssymb, amsmath}
\DeclareMathAlphabet{\mathcal}{OMS}{cmsy}{m}{n}

\usepackage{graphicx}
\usepackage{color}
\usepackage{caption}
\usepackage{subcaption}

\usepackage{multirow}
\usepackage{tabularx}
\usepackage{booktabs}

\usepackage{cite}

\usepackage{cancel}
\usepackage{framed}
\usepackage{url}
\usepackage{hyperref}
\hypersetup{colorlinks,linkcolor=black, citecolor=blue, urlcolor=blue}

\title{\LARGE \bf Depreciation Cost is a Poor Proxy for Revenue Lost to Aging in Grid Storage Optimization}

\author{Volkan Kumtepeli$^{1}$, Holger Hesse$^{2}$, Thomas Morstyn$^{1}$, Seyyed Mostafa Nosratabadi$^{1}$, Marko Aunedi$^{3}$\\ and David A. Howey$^{1}$
\thanks{\footnotesize*The authors acknowledge UKRI funding (ref.\ EP/W027321/1).}%
\thanks{\footnotesize$^{1}$Volkan Kumtepeli, Thomas Morstyn, Seyyed Mostafa Nosratabadi and David Howey are with the Department of Engineering Science, University of Oxford, OX1 3PJ, Oxford, UK
        {\tt\footnotesize \{david.howey, volkan.kumtepeli, thomas.morstyn, mostafa.nosratabadi\}@eng.ox.ac.uk}  }%
\thanks{\footnotesize$^{2}$Holger Hesse is with the Kempten University of Applied Sciences, Bahnhofstr.\ 61, 87435 Kempten, Germany
        {\tt\footnotesize holger.hesse@hs-kempten.de}}%
\thanks{\footnotesize$^{3}$Marko Aunedi is with the Department of Electronic and Electrical Engineering, Brunel University London, Uxbridge UB8 3PH, UK
        {\tt\footnotesize marko.aunedi@brunel.ac.uk}}%
}

\begin{document}
\maketitle
\thispagestyle{empty}
\pagestyle{empty}

\begin{abstract}
Dispatch of a grid energy storage system for arbitrage is typically formulated into a rolling-horizon optimization problem that includes a battery aging model within the cost function. Quantifying degradation as a depreciation cost in the objective can increase overall profits  by extending lifetime. However, depreciation is just a proxy metric for battery aging; it is used because simulating the entire system life is challenging due to computational complexity and the absence of decades of future data. In cases where the depreciation cost does not match the loss of possible future revenue, different optimal usage profiles result and this reduces overall profit significantly compared to the best case (e.g., by 30--50\%). Representing battery degradation perfectly within the rolling-horizon optimization does not resolve this---in addition, the economic cost of degradation throughout life should be carefully considered.  For energy arbitrage, optimal economic dispatch requires a trade-off between overuse, leading to high return rate but short lifetime, vs.\ underuse, leading to a long but not profitable life. We reveal the intuition behind selecting representative costs for the objective function, and propose a simple moving average filter method to estimate degradation cost. Results show that this better captures peak revenue, assuming reliable price forecasts are available. 
\end{abstract}

\section{Introduction}
\begin{figure}
    \centering
\includegraphics[width=0.9\linewidth]{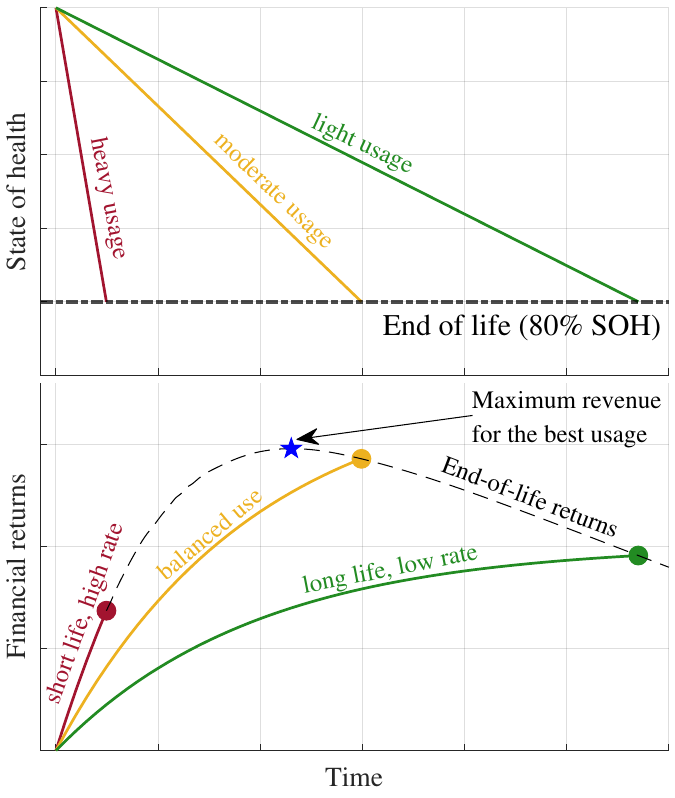}
    \caption{Maximizing the returns of grid energy storage requires a trade-off between overuse, resulting in limited revenue because of early end-of-life, and underuse, resulting in limited revenue because of missed opportunities. Dots indicate end-of-life returns for each case.}
    \label{fig:summary}
\end{figure}

Grid-scale battery energy storage systems (BESSs) are a key technology to enable zero-carbon power \cite{grunewald2023taking}, with lithium-ion batteries the popular choice due to their high energy density and flexibility. However, advanced control of charging and discharging, accounting for battery lifetime, is required to unlock the full economic potential of a BESS \cite{reniers2021unlocking}. To this end, researchers have focused on representing aging as accurately as possible within the dispatch optimization so as to minimize damaging usage, such as holding a battery at extreme temperature or state of charge (SOC). Various aging models have been used within techno-economic analyses, such as semi-empirical \cite{kumtepeli2019design, hesse2019ageing,kumtepeli2020energy, collath2023increasing}, physics-based \cite{reniers2021unlocking,vykhodtsev2023physics} or machine-learning \cite{aitio2021predicting} approaches. Studies have also shown that it is possible to squeeze more from a battery by strategically stacking different user applications, with minimal compromise on battery health by balancing different degradation modes \cite{Reniers2018ImprovingModelling,kumtepeli2022understanding, englberger2020unlocking,englberger2021electric}. Efforts towards improved BESS optimization are summarized by various reviews \cite{hesse2017lithium, Reniers2019ReviewBatteries,collath2022aging}. In essence, the usage profile has a large impact on lifetime and financial returns. As a toy example, assuming price volatility is similar throughout life, Fig.\ \ref{fig:summary} shows that adopting an aggressive dispatch strategy that capitalizes on every revenue opportunity decreases lifetime significantly, diminishing potential future revenue because the battery dies early. At the opposite extreme, a strategy that reserves battery deployment only for the highest price volatility events results in underuse, again diminishing overall revenue. Therefore, striking a balance between these extremes is important. 

For a typical BESS application, overall profit $\mathbb{P}$ to be maximized in some time period is given as \cite{hesse2017lithium,hesse2019ageing}
\begin{align}
    \mathbb{P} = \mathbb{R}^\text{app} - \mathbb{C}^\text{aging}, 
\end{align}
where $\mathbb{R}^\text{app}$ denotes the attainable revenue from the application, and $\mathbb{C}^\text{aging}$ represents the monetary `cost' of battery degradation. Generally $\mathbb{C}^\text{aging}$ is modeled as
\begin{align}
    \mathbb{C}^\text{aging} = c_\text{ag} Q_\text{loss},
\end{align}
where $Q_\text{loss}$ is the lost capacity corresponding to the history of battery usage and $c_\text{ag}$ is the corresponding financial penalty for capacity loss. The value of $c_\text{ag}$ can be determined by different approaches; for instance, it might be considered static or dynamic; it could be based on the initial investment or future value of the battery; it could even be chosen arbitrarily \cite{garcia2022review}. 

For example, Wankmüller et al.\ showed the change of net present value (NPV) by a parameter sweep of degradation penalty cost $c_\text{ag}$ with respect to two different aging models; results are not general, but hint that penalizing degradation could result in higher overall revenues\cite{wankmuller2017impact}. As an extension of their previous work \cite{he2018intertemporal}, a study by He et al.\ \cite{he2020economic} followed a more systematic approach, defining an economic end-of-life (EOL) criterion and highlighting the importance of using the future value of the revenue streams during dispatch. Recently, Collath et al.\ \cite{collath2023increasing} conducted a detailed analysis of a battery within a German market to quantify the economic advantages. They reported the benefit of adapting the aging factor through the years, considering the time value of money (i.e., the interest rate). Despite these studies emphasizing the importance of selecting the optimal degradation cost coefficient, many researchers (including us, e.g.\ \cite{kumtepeli2020energy, reniers2021unlocking}) have focused only on adding the most accurate degradation information into the optimization as a static depreciation cost; further analysis has been limited to simple parameter sweep/sensitivity analysis or very specific case studies.

In this study, we generalize the findings of He et al.\ \cite{he2020economic} and Collath et al.\ \cite{collath2023increasing} through further investigation of the cost function used in storage optimization. The overall aim is to establish the beginnings of a more systematic approach for determining battery cost within the real-time optimizer.

\section{Methodology}

We focus on grid BESS energy arbitrage as the application since this allows us to freely scale revenue upwards (with a saturating behavior -- limited by the battery capacity) while having to account for additional aging-related burdens (i.e.\ increasing revenue comes at the cost of additional aging). To test the effect of giving `exact' degradation information to the optimizer, we assumed simplified time-independent and convex versions of semi-empirical degradation models \cite{kumtepeli2020energy}, along with the linear  dynamic model of Hesse et al.\ \cite{hesse2019ageing} and constant BESS efficiency. Hence, the model can be exactly represented in the cost function without approximation errors. The weights of the aging functions were then adjusted over a wide range. The final objective function is given by
\begin{align}
    \label{eq:perturbed_objective}
    J = \sum_{t \in \mathcal{T}} \left( c_\text{en} P_\text{AC} \Delta t + c_\text{ag} \left( \lambda_\text{cyc}Q_\text{loss-cyc} + \lambda_\text{cal} Q_\text{loss-cal}\right) \right),
\end{align}
where $c_\text{en}$ represents the intraday electricity market price at each timestep $t$, $P_\text{AC}$ is a sequence of uniformly spaced (dis)charge powers to or from the battery, $\lambda_\text{cyc}$ and $\lambda_\text{cal}$ are the aging function weights, and $Q_{\text{loss}}$ is a capacity fade, both in each case given respectively for cycle and calendar aging. The degradation cost penalty $c_\text{ag}$ was defined as
\begin{align}
    \label{eq:typical_selection}
    c_\text{ag} = c_\text{battery}/Q_\text{EOL},
\end{align}
where $c_\text{battery}$ is the initial battery investment cost and $Q_\text{EOL}$ denotes the fractional amount of capacity loss by end-of-life (EOL), which we here assume to be 20\%, equating to 80\% of the original capacity---this is a commonly accepted economic end-of-life state of health (SOH) fraction \cite{collath2022aging, kumtepeli2019design}. We can think of $c_\text{ag}$ as a cost per unit of capacity fade---for example, if we pay \$1000 for a battery, then $c_\text{ag}=$ \$50 per cent of capacity fade (if $Q$ is expressed as a percentage). If we lose 10\% of the capacity, we have therefore lost \$500 value, which is half of the initial cost of the battery (under the assumption that EOL is at 80\% capacity). 

The objective function is constrained by the battery dynamics, \eqref{eq:Pbatt}--\eqref{eq:loss}, where the superscripts $+/-$ denote charging/discharging respectively. We model dynamics in discrete time and assume a zero-order hold between steps. Power is related to C-rate by  
\begin{align}
\label{eq:Pbatt}
0 \leq P_\text{batt}^{+/-}(t) &\leq E_\text{nom} C_\text{r-max}^{+/-},   
\end{align}
where $P_\text{batt}$ and $E_\text{nom}$ are the power and nominal beginning of life (energy) capacity, and $C_\text{r-max}^{+/-}$ are upper limits on charging and discharging C-rates. (Note, this may alternatively be defined using the reciprocal of C-rate, i.e., duration.) Battery power $P_\text{batt}$ is linked to AC-side power $P_\text{AC}$ by
\begin{align}
\label{eq:AC_side}
P_\text{AC}(t)&= P_\text{batt}^+(t)/\eta^+ - P_\text{batt}^-(t)  \eta^- ,
\end{align}
where $\eta^{+/-}$ are charge/discharge efficiencies, lumping both battery and power electronics. The dynamics for the stored energy are represented as follows: 
\begin{align}
\label{eq:ebattery1}
E_\text{batt}(0) &=  E_0 \\
E_\text{batt}(t) &= E_\text{batt}(t-1) + (P_\text{batt}^+(t) - P_\text{batt}^-(t)) \Delta t   \\
0 \leq E_\text{batt}(t) &\leq E_\text{nom} \text{SOH}_0
\end{align}
\begin{figure*}
  \centering
  \subfloat[Interest rate $i=0\%$ \label{fig:perturbation}]{%
       \includegraphics[height=0.3247\linewidth]{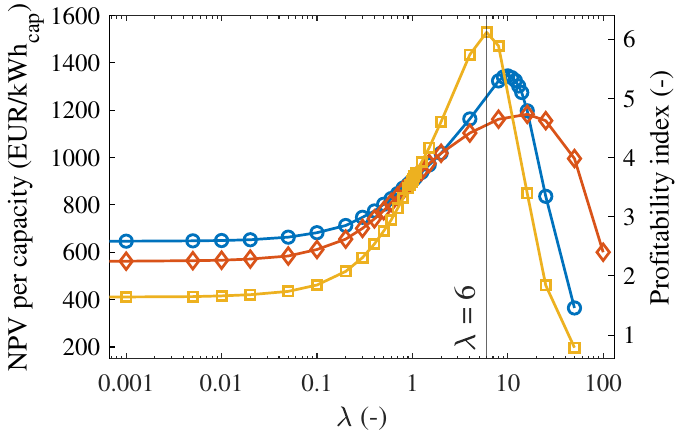}}\hspace{1pt}
  \subfloat[Interest rate $i=5\%$ \label{fig:profitabilityindex}]{%
        \includegraphics[height=0.3247\linewidth]{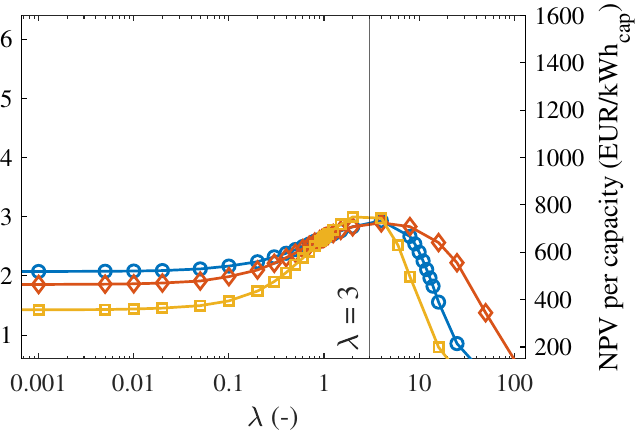}}
  \caption{Net present values and corresponding profitability indices for different degradation cost weights and interest rates; firstly where only $\lambda_\text{cal}$ is changed and $\lambda_\text{cyc}=1$ (blue line with circle markers); secondly where only $\lambda_\text{cyc}$ is changed and $\lambda_\text{cal}=1$ (red line with diamond markers); thirdly, with $\lambda_\text{both} = \lambda_\text{cal} = \lambda_\text{cyc}$ and both are changed together (yellow line with square markers).}
  \label{fig:Revenue_NPV_sweep} 
\end{figure*}
Here, $E_\text{batt}$ is the usable energy in the battery at each time step, with initial value  $E_\text{0}$ at the start of each rolling window, constrained by a slowly diminishing upper limit given by the beginning of life energy $E_\text{nom}$ multiplied by a quasi-steady but gradually decreasing state of health factor, SOH$_0$, which is fixed for an individual optimization run, but updated each day according the rolling horizon scheme used. To model thermal effects we use the approach of Kumtepeli et al.\  \cite{kumtepeli2020energy},
\begin{align}
T(0) &=  T_0 \\
T(t) &= T(t-1) + k_\text{T}\left(\alpha_\text{T}(T_\text{amb} - T(t-1) + \dot{Q}_\text{T}\right)\Delta t \\
\dot{Q}_\text{T}(t) &=  (\beta^+P_\text{batt}^+ + \beta^-P_\text{batt}^-(t))/{E_\text{nom}} \\ 
T_\text{avg}(t) &= \left(T(t-1) + T(t)\right)/2,
\end{align}
where $T$, $T_\text{avg}$ and $T_\text{amb}$ are respectively the instantaneous and average battery and ambient temperatures, $\dot{Q}_\text{T}$ is the heat generation rate, and $k_\text{T}$, $\alpha_\text{T}$, $\beta^+$, $\beta^-$ are fitted constants that depend on the system. 

Before defining the aging model, we define the change in full equivalent cycle count at each time step, $\Delta \text{FEC}$, and the average SOC, as 
\begin{align}
\Delta \text{FEC}(t) &= \frac{ P_\text{batt}^+(t)+ P_\text{batt}^-(t)}{2 E_\text{nom}}\Delta t \\
\text{SOC}_\text{avg}(t)&= \frac{ E_\text{batt}(t-1)+  E_\text{batt}(t)}{2 E_\text{nom} \text{SOH}_0} .
\end{align}

Given the superposition principle, the effects of calendar and cycle aging factors are represented \cite{hesse2017lithium} by decoupled models \eqref{eq:aging_first}--\eqref{eq:loss}; either linearly, for cycle aging during discharge, or with  piecewise-affine approximation---for cycle aging during charging, and for calendar aging. These were selected to ensure convexity and avoid binary variables; they contain multiple planes that are indexed by $i$,   $\forall i \in \mathcal{I}_\text{cyc}$ for cycle aging $Q_\text{loss-cyc}$, and $\forall i \in \mathcal{I}_\text{cal}$ for calendar aging $Q_\text{loss-cal}$.
\begin{align}
Q_\text{loss-cyc}^-(t) &= \text{k}_\text{loss-cyc}^-\Delta \text{FEC}(t)  \label{eq:aging_first} \\
Q_\text{loss-cyc}^+(t) &\geq a_\text{cyc}(i) + b_\text{cyc}(i) P_\text{batt}^+(t)/ E_\text{nom}  \\ 
Q_\text{loss-cyc}(t) &= Q_\text{loss-cyc}^-(t) + Q_\text{loss-cyc}^+(t) \\ 
Q_\text{loss-cal}(t) &\geq a_\text{cal}(i) +  b_\text{cal}(i) \text{SOC}_\text{avg}(t) + c_\text{cal}(i)T_\text{avg}(t) \label{eq:loss}
\end{align}

The optimization problem, \eqref{eq:perturbed_objective}-\eqref{eq:loss}, was implemented in Python via CVXPY and Gurobi \cite{diamond2016cvxpy, agrawal2018rewriting,gurobi}, and optimizations were run until the assumed EOL condition of 80\% SOH, using a 7-day window, with 1-day overlapping rolling horizon, and a 15 min time step discretization size. The variables $E_0$, $\text{SOH}_0$, and $\text{T}_0$ were updated after each step. One year of price information was obtained from the public dataset in our former work \cite{kumtepeli2020energy}; this was transformed into a nonnegative time series by taking the absolute values of the prices to avoid breaking the convexity of the piecewise models and to avoid causing concurrent charging and discharging.

\section{Results and discussion}
The parameter $\lambda$ was swept over a wide range of values, with the optimization problem solved at each value. The resulting financial performance was evaluated by considering the net present value (NPV) of the total revenue over the entire battery lifetime \eqref{eq:NPV}, as well as a \emph{profitability index}, PI \cite{kumtepeli2020energy}, i.e., NPV divided by battery upfront cost:
\begin{align}
    \label{eq:NPV}
    \mathbb{R}^\text{NPV} &= \sum_{p=1}^{\lceil t_\text{EOL} \rceil} \left(\frac{\mathbb{R}^\text{app}_p}{(1+i)^p}\right), \\ 
    \text{PI} &= \mathbb{R}^\text{NPV} / c_\text{battery}. \label{eq:PI}
\end{align}
Here, $i$ is the annual interest rate, and  $\mathbb{R}^\text{app}_p$ is the $p$-th year's revenue. The results are presented in Fig.\ \ref{fig:perturbation} with $i=0\%$ (i.e., ignoring the time value of money), and in Fig.\ \ref{fig:profitabilityindex} assuming an interest rate $i=5\%$. This highlights that total attainable revenue is not necessarily maximized by providing the optimizer with perfect degradation information (i.e., both $\lambda=1$, an exact representation of degradation in the cost function) and a typical degradation cost based on initial investment cost (i.e., \eqref{eq:typical_selection}). Instead, the peak revenue is obtained at different $(\lambda_\text{cyc}, \lambda_\text{cal})$ pairs of $(1,10)$, $(16,1)$, or $(6,6)$ for the different parameter sweeps as shown in Fig.\ \ref{fig:perturbation}; equivalently in Fig.\ \ref{fig:profitabilityindex} the optimal joint value is $(3,3)$. (Note that nothing has been changed in the forward simulation; only the $\lambda$ values have been changed.) Moreover, different aging components ($Q_\text{loss-cal}$ and $Q_\text{loss-cyc}$) reach a peak revenue at different $\lambda$ values. This also shows the importance of balancing between these aging effects. The situation considering 5\% interest rate is similar to the one at zero interest but with a smaller profitability, reached at smaller $\lambda$ values. Hence, we observe that the initial simple investment-based degradation cost does not align with potential future profits, suggesting the need for a revised degradation penalty based on predicting the future revenue.

Although $\lambda_\text{cal}$ and $\lambda_\text{cyc}$ reach peak revenues at different values under individual parameter sweep analyses (where we keep one of them constant), a two-dimensional parameter sweep analysis reveals that, given our model, their values are closely related---as shown in Fig.\ \ref{fig:appendix-sweep2d} in the Appendix. Therefore, we simplify the analysis from here by perturbing both calendar and cycling $\lambda$ values at the same time, defining $\lambda_\text{both} = \lambda_\text{cal} = \lambda_\text{cyc}$. Further information regarding the behavior of calendar and cycle aging for different parameter sweeps (Fig.\ \ref{fig:appendix-1}) can be found in the Appendix.

To find the optimal combined $\lambda_\text{both}$ value, we begin by revisiting the primary objective---to maximize profitability over the battery lifespan, essentially optimizing revenue per unit of battery degradation (quantified as a capacity fade percentage in a given amount of time). Aggressive exploitation of a battery for marginally higher revenue leads to decreased profitability per amount of capacity fade and jeopardizes the opportunity to obtain future revenue. Thus, conserving the battery health for higher-value future opportunities is important. The critical question then is how to discern, at each moment during real-time operation, which opportunities merit battery dispatch. Understanding the expected revenue of the target operation is vital. We hypothesize that the cost associated with aging should be contingent upon two factors,
\begin{enumerate}
    \item the battery use today (which causes a decrease in future remaining useful life), and,
    \item the anticipated value/cost of that degradation in the future (i.e., revenue per amount of capacity fade, or weight associated with $Q_\text{loss}$).  
\end{enumerate}
\begin{figure}
    \centering
    \includegraphics[width=\linewidth]{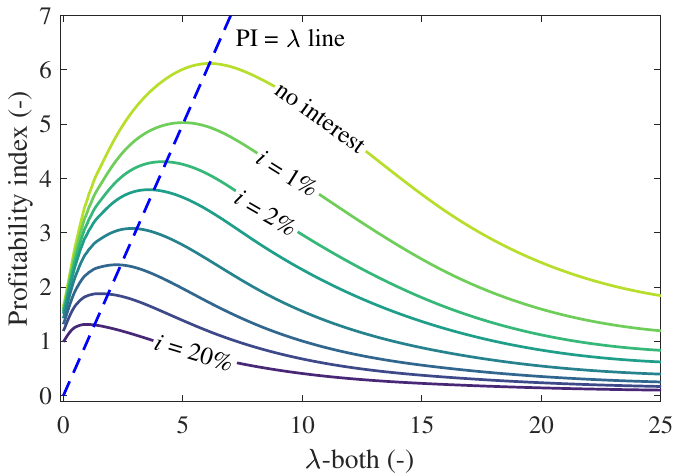}
    \caption{Change in profitability index (PI) as a function of $\lambda_\text{both}$ for various interest rates. Solid lines represent yearly interest rates of 0\%, 1\%, 2\%, 3\%, 5\%, 8\%, 12\%, and 20\% respectively, top to bottom. Dashed line is PI = $\lambda_\text{both}$. }
    \label{fig:CC_vs_PI}
\end{figure}
This was investigated using the extensive simulation data obtained already from the parameter sweep. Since the simulations quantify the total revenue gained until  EOL, we hypothesize the optimal value of $\lambda$ to be
\begin{align}
    \lambda \approx \frac{\mathbb{R}^\text{NPV}/ Q_\text{EOL}}{c_\text{ag}} = \frac{\mathbb{R}^\text{NPV}/ Q_\text{EOL}}{c_\text{battery}/Q_\text{EOL}}= \frac{\mathbb{R}^\text{NPV}}{c_\text{battery}} = \text{PI}.
\end{align}
In other words, the optimal $\lambda$ value is approximately equal to net present value of the total revenue over all of life divided by the initial investment cost, i.e., the profitability index \eqref{eq:PI}. The correspondence between optimal $\lambda$ and PI can be observed in Fig.\ \ref{fig:Revenue_NPV_sweep}---in Fig.\ \ref{fig:perturbation}, the maximum profit is attained at $\lambda_\text{both}=6$ which coincides with $\text{PI} \sim 6$. Similarly, in Fig.\ \ref{fig:profitabilityindex}, although they do not perfectly match, the peak value of the profitability index is $\sim$3, and the corresponding $\lambda_\text{both}$ value resides around 3. Further insight into this issue can be found in Fig.\ \ref{fig:CC_vs_PI}, which plots the profitability index versus $\lambda$ for different interest rates. %
 
Although our hypothesis of $\lambda \approx \text{PI}$ gives a high-quality guess for optimal $\lambda_\text{both}$, there is a small mismatch between the peak profit and the line $\text{PI} = \lambda_\text{both}$ (indicated on the figure) when interest rate increases. This, we assume, supports the case for a time-varying $\lambda$; however, an alternative method for a slight improvement in the selection of optimal $\lambda_\text{both}$ is speculatively given in Fig.\ \ref{fig:appendix-optLambdaCorrection} in the Appendix. 
\begin{figure}
    \centering
\includegraphics[width=\linewidth]{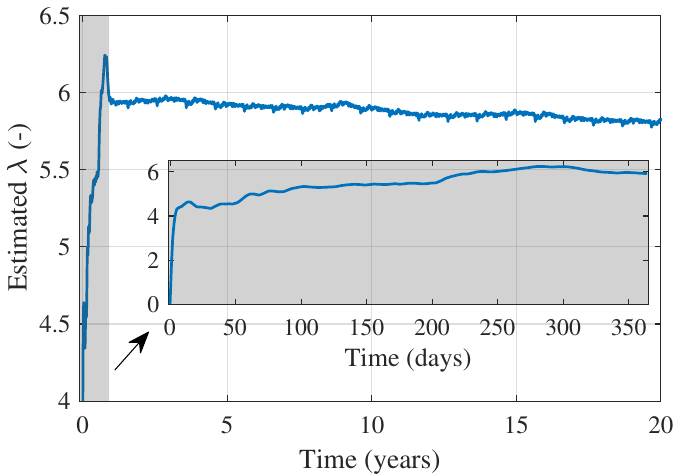}
    \caption{Evolution of estimated $\lambda$ using moving mean over historical data, with a window size of one year. Zoomed section (grey) shows performance in the first year.}
    \label{fig:lambda_evo}
\end{figure}

In practice, it is not possible to run simulations over the entire battery lifetime, and it is unrealistic to assume, e.g., that electricity price volatility is fixed throughout life. We therefore now investigate whether it is possible to approximate and track the optimal value of $\lambda_\text{both}$ that yields the peak revenue, without having to undertake extensive whole-life simulations. We show that this is possible as long as one can forecast the future profit of the system. To find the expected value of the revenue/aging ratio, one can use a simple moving-average filter with recent historical data as the input. In this approach, after each rolling-horizon optimization step, we estimate the amount of aging incurred and revenue earned via forward simulation; then we record their ratios in an array representing the moving-horizon window. Lastly, we update $\lambda_\text{both}$ at each step by calculating the mean of the values in the array. After trying different window lengths (1 week to 1 year) and finding out that they perform similarly (but longer horizons performing better), we present results regarding the evolution of $\lambda_\text{both}$ in Fig.\ \ref{fig:lambda_evo} for a one-year window length. This simple approach performs surprisingly well and is able to find and track a reasonable $\lambda_\text{both}$ value within a week. We can also see the decline in the value of $\lambda_\text{both}$ due to the capacity loss caused by degradation. The moving-average method achieves a profit of 1528.9 EUR/kWh$_\text{cap}$ which is very close to the peak value of 1529.3 EUR/kWh$_\text{cap}$ found by the parameter sweep. This method is capable of tracking the optimum value, but with large caveats: if (and only if) recent historical revenue data (within the window of the moving average filter) is representative of future possible revenue data, and aging is slow and smooth. Despite this, using an adaptive penalty for degradation cost could result in much better performance than using a constant or fixed cost throughout life, although further work is required to pressure test this idea.   
\begin{figure}
    \centering
\includegraphics[width=\linewidth]{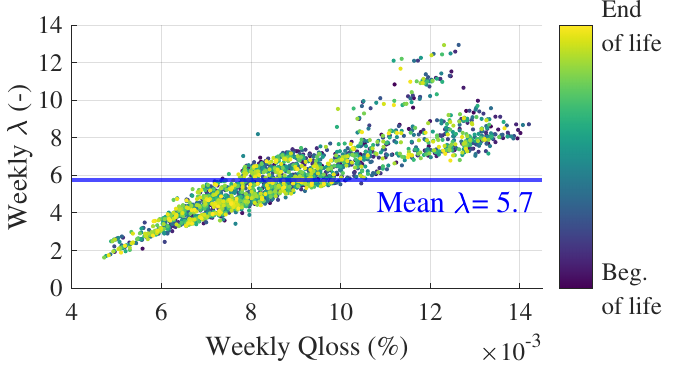}
    \caption{Estimated $\lambda$ based on weekly revenue and $Q_\text{loss}$, colors denote the closeness of the data point to end of life.}
    \label{fig:LS}
\end{figure}
\section{Conclusions}
In this paper, we investigated the importance of selecting a suitable cost coefficient for the aging term in a battery economic optimization problem to represent the future profit loss caused by degradation. The results reveal the simple yet interesting connection between the weighting of degradation cost and the optimal total profitability over lifetime of the battery. Summarizing, it may be economically beneficial to optimize `revenue per unit of degradation' in grid storage systems, rather than 'revenue minus depreciation cost'. This is just a starting point, and further investigation on this topic is required. To give a flavor of one possible direction, Fig.\ \ref{fig:LS} shows the estimated $\lambda$ values given the weekly revenue and degradation, ignoring discounting. Although the mean $\lambda$ is very close to the optimal value of 6, it actually lower ($\sim$5.7) which could be attributed to outliers. It could be argued that these outliers are actually the ones we need to capture, and therefore, more advanced estimation techniques should be used to infer $\lambda$. Moreover, it may be possible to use state-of-the-art machine learning techniques to forecast future revenue and embed the resulting information into $\lambda$. More accurate aging models could also reveal more information about the difference between calendar and cycle aging costs. Finally, future work could explore an adaptive solution \cite{perriment2023leadacid} for changing conditions battery \cite{he2020economic}. One such condition is the nonlinear battery aging that occurs past a `knee point' where capacity fade accelerates.

\begin{figure}
    \centering
\includegraphics[width=1\linewidth]{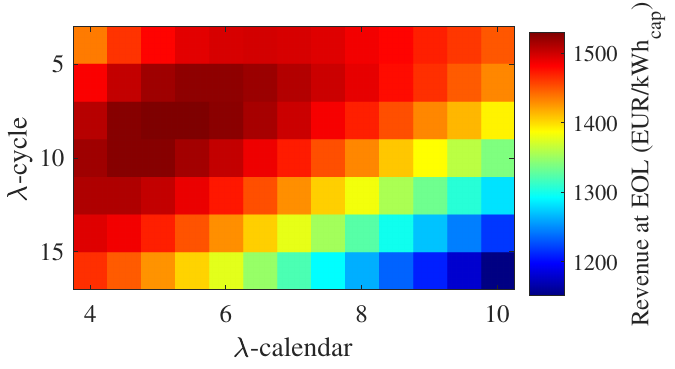}
    \caption{Overall revenue for different degradation cost weights where both $\lambda_\text{cyc}$ and $\lambda_\text{cal}$ are changed independently.}
    \label{fig:appendix-sweep2d}
\end{figure}

\appendix
Fig.\ \ref{fig:appendix-sweep2d} shows the result of a two dimensional parameter sweep where the best $\lambda$ values are similar to the case where they are changed simultaneously (i.e.\ $\lambda_\text{both} = \lambda_\text{cal} = \lambda_\text{cyc}$).

\begin{figure*}
    \centering
  \subfloat[Perturbation of $\lambda_\text{cal}$ \label{fig:appendix-1a}]{%
       \includegraphics[width=0.33\linewidth]{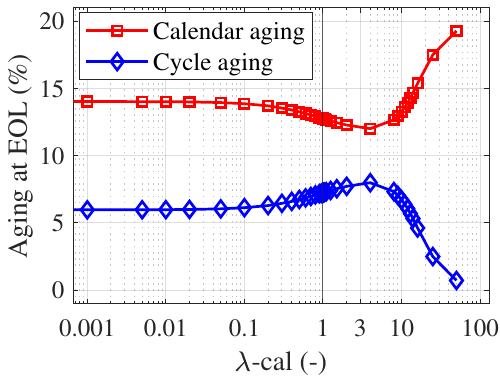}}
  \subfloat[Perturbation of $\lambda_\text{cyc}$ \label{fig:appendix-1b}]{%
        \includegraphics[width=0.33\linewidth]{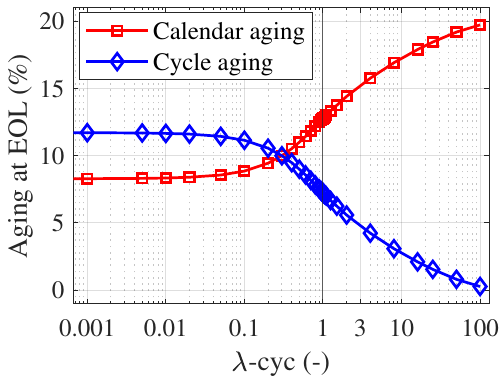}}
  \subfloat[Perturbation of both $\lambda_\text{cal}$ 
 and $\lambda_\text{cyc}$ \label{fig:appendix-1c}]{%
        \includegraphics[width=0.33\linewidth]{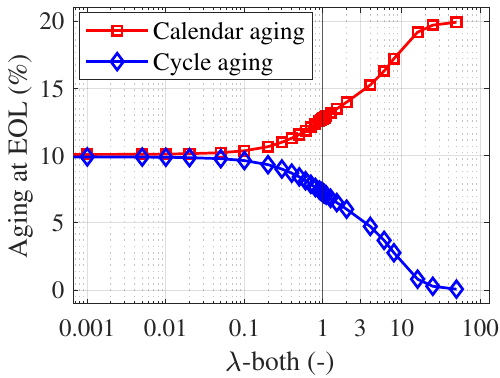}}

  \caption{(a)-(c) show how aging portions change under different perturbations.}
  \label{fig:appendix-1} 
\end{figure*}

In Fig.\ \ref{fig:appendix-1} the difference between the calendar aging and cycle aging portions of EOL capacity fade under different parameter sweeps is shown. This demonstrates that the simplifications in the aging function, unfortunately, may be poorly represented because calendar aging mostly dominates. 

\begin{figure}
    \centering
\includegraphics[width=1\linewidth]{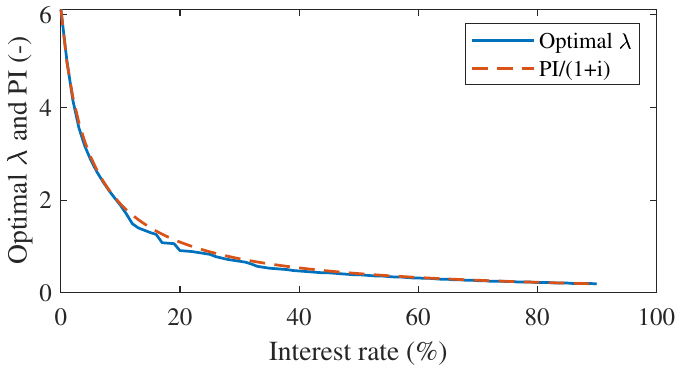}
    \caption{Alignment of optimal $\lambda$ and PI for a wide range of interest rate values, after dividing PI with a correction factor of $(1+i)$ where $i$ is the interest rate. This shows that the mismatch between optimal $\lambda$ and PI can be remedied by a correction factor.}
    \label{fig:appendix-optLambdaCorrection}
\end{figure}

Fig.\ \ref{fig:appendix-optLambdaCorrection} speculates regarding a correction factor to remedy the mismatch between optimal $\lambda$ and profitability index. 

Code for the study in this paper is available at\\
\href{https://github.com/Battery-Intelligence-Lab/BatteryOpportunity/}{github.com/Battery-Intelligence-Lab/BatteryOpportunity}
\hypersetup{urlcolor=black}

\small
\bibliography{references}
\bibliographystyle{IEEEtran}

\end{document}